\documentclass[aps,prd,twocolumn,nofootinbib]{revtex4-2}
\usepackage{epsfig}
\usepackage{graphicx}
\usepackage{dcolumn}
\usepackage{amssymb,amsmath}
\usepackage{mathrsfs}
\begin{document}

\title{Isotope shift for total electron binding energy of atoms}

\author{V. A. Dzuba}\email{v.dzuba@unsw.edu.au}
\author{V. V. Flambaum}\email{v.flambaum@unsw.edu.au}

\affiliation{School of Physics, University of New South Wales, Sydney 2052, Australia}

\begin{abstract}
We compute the isotope shifts of the \emph{total}  electron binding energy of neutral atoms and singly charged  ions up to element $Z=120$, using  relativistic Hartree-Fock method including the Breit interaction.
Field shift coefficients are extracted by varying the nuclear charge radius; a small quadratic term is retained to cover large radius changes relevant to superheavy nuclei.
We tabulate isotope shift coefficients for closed shell systems from Ne to Og and benchmark selected open shell cases, used to test the interpolation formula.
A simple power law interpolation  $bZ^k$ reproduces calculated field shifts to within about 1\% across the table, with the effective exponent $k$ growing from roughly 5 near $Z \sim 50$ to about 12 at $Z \sim 118$.
Due to the domination of inner shells, differences between neutrals and singly charged ions does not exceed few percent, becoming noticeable mainly when an outer $s$ electron is removed. Therefore, these results may also be used for  higher charge ions. 
\end{abstract}

\maketitle

\section{Introduction}

Study of the total electron binding energy began soon after the advent of quantum mechanics. Beyond intrinsic atomic physics interest, this quantity is essential for precise nuclear mass accounting and for testing nuclear models.  Nuclear masses are measured  with relative mass precision  $\delta M/M \sim  10^{-10}$   \cite{MLO.03,DBBE.18}. To extract bare nuclear masses from measured atomic masses, one must subtract the total electron binding energy with comparable accuracy.

Accurate nuclear binding energies underpin determinations of the limits of nuclear stability and the development of models of nucleon interactions in nuclear matter (see e.g. \cite{MLO.03,Eet.12,AARR.13,AARR.14,PRC}). Such interactions govern properties of neutron stars \cite{CH.08} and the synthesis of heavy elements \cite{TEPW.17}. They also inform predictions of neutron distributions (neutron skins), which influence atomic parity violation observables used to test the Standard Model and probe new physics \cite{Brown,Antypas,Stadnik,Roberts}.

For neutrino mass experiments, selected isotope mass differences must be known to better than 1 eV \cite{DBBE.18}. Precise nuclear masses are likewise critical for searches for new scalar mediated interactions using high accuracy isotope shift spectroscopy \cite{Berengut}. Finally, many nuclear astrophysical processes depend sensitively on nuclear masses because these enter reaction energy release   and rates \cite{AGT.07}.

First calculations of the total electron binding energy were performed using the Thomas- Fermi method ( see e.g. book \cite{Landau}). Later  a number of accurate calculations were  performed (see e.g.  \cite{HACCM.76,Santos,fit}). The most complete calculations have been performed in our paper  ~\cite{Etot}  for all atoms up to $Z=120$. The  relativistic Hartree-Fock (RHF) method  was used to find a zero approximation. Then the correlation, Breit and quantum electrodynamics (QED) corrections have been calculated. For the convenience of use, all results have been fitted by simple  analytical  functions of the nuclear charge $Z$.

However, the isotope shift (IS) - the difference in total electron binding energy between isotopes - was not considered in Ref.~\cite{Etot}. In applications such as determining the boundaries of nuclear stability (locating  the proton and neutron drip lines), the differences in mass number $A$ among isotopes can be large. There is a substantial gap between the superheavy isotopes produced in laboratories and those associated with the hypothetical island of stability. Some elements  have many stable isotopes. For example,  Sn has ten stable isotopes with atomic number $A$ ranges from 112 to 124.
Searches for new scalar mediated interactions using high accuracy isotope shift spectroscopy are very sensitive to the differences of the isotope masses \cite{Berengut}.
 Precise isotope \emph{mass differences} are central to isotope separation techniques and can often be determined more accurately than absolute nuclear masses, which motivates a dedicated evaluation of the IS contribution to these differences.

The principal correction to the simple  sum of the added neutron masses arises from nucleon - nucleon interactions within the nucleus. Variations in the total electron binding energy likewise contribute to atomic isotope mass differences. This is an atomic structure problem that warrants attention in its own right, independent of nuclear physics applications. In Ref.~\cite{PRC} we computed the isotope shift of the total electron binding energy for several heavy atoms; to the best of our knowledge, no prior calculations of this quantity exist.

In the present work we carry out a comprehensive study of the isotope shift (IS) in the total electron binding energy for all neutral atoms and singly charged ions with $Z\le 120$. The difference in IS between neutrals and singly charged ions is small, because the dominant contribution to the total-energy IS comes from deep inner-shell electrons. A few percent difference appears only when an $s$ electron is removed. Therefore, our results can also be used for ions with higher charge states.

The calculations are performed with the relativistic Hartree-Fock (RHF) method including the  Breit interaction. 
%We consider both the mass shift and the field shift.
 Most calculations are for closed shell systems; for open-shell atoms we provide an accurate interpolation formula between neighboring closed shells atoms. To test this formula, we performed calculations for a number of  open-shell  elements.

\section{Calculations}

The difference in total electron binding energy between two isotopes can be approximated by (see, e.g. \cite{Sobelman})
\begin{eqnarray}
&&\Delta E_{\rm IS} = F\delta\langle r^2\rangle + G\delta\langle r^2\rangle^2 + \label{e:FG}\\
&&\left(\frac{m_e}{M_1}-\frac{m_e}{M_2}\right)\left(K_{\rm NMS}+K_{\rm SMS}\right). \label{e:MS}
\end{eqnarray} 
Here the first line represents the field shift (FS) (shift in energy due to change of  the squared nuclear radius $\delta\langle r^2\rangle$ ), the second line represents the mass shift (MS),
which consists of two contributions, the normal mass shift (NMS) and the specific mass shift (SMS), $m_e$ is the electron mass. 
MS dominates in light atoms ($Z \lessapprox 40$, see below), FS dominates in heavy atoms~\cite{Sobelman}.
The term with $G$ is often neglected due to its small contribution. We keep it to cover the case of large $\delta\langle r^2\rangle$, e.g.
between heaviest observed isotope of a heavy element and hypothetical isotope in the island of stability. For example, the heaviest observed 
isotope of nobelium is $^{259}$No. It has 157 neutrons, while a more stable isotope should have 184 neutrons. This large change in atomic number
($\Delta A =27$) should lead to large  $\delta\langle r^2\rangle$.

Another reason for retaining the term with $G$ is to improve accuracy for heavy atoms.  The field shift in atomic transitions, involving $s_{1/2}$ or $p_{1/2}$ electrons, is more accurately approximated by (see e.g. Refs.~\cite{Sobelman,gamma})
\begin{equation}\label{e:2gamma}
\Delta E_{\rm FS} = F\delta\langle r^{2\gamma}\rangle, 
\end{equation}
where $\gamma = \sqrt{1-Z^2\alpha^2}$.
However,  in the case of  IS in the total electron energy this formula may be  not sufficiently accurate, and retaining the term with $G$ provides a better alternative.

We calculate field shift constants $F$ and $G$ by calculating total electron binding energy of an atom at three different values of nuclear radius.
The case of smallest radius is considered as a reference. The change of the total electron energy for other isotopes is approximated by a parabola (\ref{e:FG}).

Note that due to smooth dependence of the energy on $Z$ we perform calculations for closed-shell atoms only and approximate the results for other atoms by simple analytical interpolating formula (see below). Note that we consider external $s^2$ shell as a closed shell.
%We also included an open-shell atom Sr because of its importance as an atoms having ten stable isotopes with atomic number $A$ ranges from 112 to 124.

The normal mass shift is proportional to the total kinetic energy of all electrons, $E_k$. In the nonrelativistic limit the virial theorem gives  $E_k = -E_{\rm total}$. Since total electron binding energies were computed in Ref.~\cite{Etot}, no additional calculation of $E_k = -E_{\rm total}$ is required. The mass shift is important in light atoms only, so relativistic corrections are unnecessary.

%The normal mass shift  is proportional to the  total kinetic energy of all atomic electrons($E_k$). In the non-relativistic limit one can use virial theorem to obtain $E_k = -E_{\rm total}$.
%Since total electron energies were calculated in Ref.~\cite{Etot},  no new calculations of $E_k$  are needed. Mass shift  dominates in light atoms, therefore there is no need to consider relativistic case. 

The specific mass shift  is proportional to $K_{\rm SMS} = 2\sum_{i<j} \langle \mathbf{p}_i \cdot \mathbf{p}_j\rangle.$ 
No simple approximation exists for this term, so it must be evaluated explicitly. In transition frequencies the SMS is extremely sensitive to electron-correlation effects, making accurate calculations challenging  (see, e.g. ~\cite{SMS1,SMS2}). For the total binding energy, however, the dominant contribution comes from the innermost shells, where correlations are small. For example, in xenon about 40\% of the total energy SMS arises from the $1s$ electrons, and 80\% from the $1s$, $2s$ and $2p$ shells combined. We therefore evaluate  SMS as an expectation value of its operator in the relativistic Hartree-Fock  ground state.

Our results show that the SMS is a modest correction to the NMS, growing from roughly 10\%  in light atoms to slightly above 20\%  in Xe. Given that the ratio of  the  mass shift to the field  shift rapidly decreases with  $Z$   (see  Table \ref{t:MS-FS}), the SMS contribution to the isotope shift of the total binding energy is not important for practical purposes.

%There is no simple approximation for this  contribution,thus it has  to be calculated. Calculation of the SMS contribution to the frequencies of atomic transitions is very sensitive to correlations and obtaining accurate and reliable results is challenging (see, e.g. ~\cite{SMS1,SMS2}). Fortunately, the situation is different for the total electron binding energies.
%Here the dominant contribution comes from inner-most electrons where correlations are not important. For example, 40\% of the SMS for the total electron binding energy of Xe comes from the $1s$ electrons, while 80\% come from the $1s,2s$ and $2p$ electrons. Therefore, we calculate SMS as an expectation value of its operator over atomic  relativistic Hartree-Fock wave function. 

%We found that SMS for the total electron binding energy is just a  fraction of NMS. Its relative value increases with $Z$ from $\sim$ 10\% for light atoms to more than 20\% for Xe. Since the total mass shift decreases with $Z$ (see Eq. (\ref{e:MS}) )and note that $M \approx A$~amu, and $A \geq 2Z$) we conclude that the SMS contribution to the isotope shift of the total electron binding energy can be neglected.

The results of the calculations for light closed-shell atoms are presented in Table~\ref{t:FGMS}. The last column of the table presents total electron kinetic energy obtained from the total electron binding energy~\cite{Etot} using the virial theorem.

\begin{table}
\caption{\label{t:FGMS} Field shift constants ($F,G$, see formula (\ref{e:FG})) and total kinetic energy $E_k$ for light atoms.
Note that $E_k = K_{\rm NMS}$; in the non-relativistic limit $E_k = - E_{\rm total}$. The values for $E_{\rm total}$ are taken from \cite{Etot}.
Abréviation GS stands for ground state.}
\begin{ruledtabular}
\begin{tabular}   {rl lccr}
\multicolumn{1}{c}{$Z$}&
\multicolumn{1}{c}{Atom}&
\multicolumn{1}{c}{GS}&
\multicolumn{1}{c}{$F$}&
\multicolumn{1}{c}{$G$}&
\multicolumn{1}{c}{$E_k$}\\
&&\multicolumn{1}{c}{conf.}&
\multicolumn{1}{c}{a.u./fm$^2$}&
\multicolumn{1}{c}{a.u./fm$^4$}&
\multicolumn{1}{c}{a.u.}\\
\hline %       a.u/fm^2     a.u./fm^4   a.u.
  10 & Ne  &$1s^22s^22p^6$ &  4.86[-6] &  -1.57[-8] &  129 \\
  12 & Mg  & [Ne]$2s^2$ &  1.04[-5] &  -1.63[-9] &  200 \\
  18 & Ar  & [Ne]$3s^{2}3p^6$ &  5.81[-5] &  -3.19[-8] &  259 \\
  20 & Ca & [Ar]$4s^2$  &  8.94[-5] &   7.39[-7] &  680 \\
  30 & Zn  & [Ar]$3d^{10}4s^2$ &  5.47[-4] &   1.90[-6] &  1794 \\
  36 & Kr  & [Zn]$4p^6$ &  1.30[-3] &  -1.12[-6] &  2788 \\
  38 & Sr  & [Kr]$5s^2$ &  1.69[-3] &  -1.88[-6] &  3177 \\
  48 & Cd  & [Kr]$4d^{10}5s^2$ &  5.54[-3] &  -8.49[-6] &  5589 \\
  50 & Sn  & [Cd]$5p^2$ &  6.73[-3] &  -8.71[-6] &  6172 \\
%  54 & Xe   &  0.0105   &  -1.88[-5] &  7440 \\
% 56 & Ba   &  0.0126   &  -2.56[-5] &  8128 \\
%  70 & Yb   &  0.0493   &  -0.0001   &  14050 \\
%  78 & Pt   &  0.1068   &  -0.0003   &  18407 \\
 % 80 & Hg   &  0.1246   &  -0.0004   &  19617 \\
%  86 & Rn   &  0.2157   &  -0.0008   &  23559 \\
% 102 & No   &  0.9679   &  -0.0050   &  36658 \\
% 118 & Og   &  4.6288   &  -0.0330   &  54662 \\
% 120 & E120 &  5.7224   &  -0.0424   &  57367 \\
\end{tabular}
\end{ruledtabular}
\end{table}

It is instructive to compare mass shift to field shift for a wide range of atoms from small nuclear charge $Z$ to the end of the periodic table.
This indicates which of the shifts needs to be calculated accurately. We performed this calculations for closed-shell atoms from $Z$=10, to $Z$=120.
The results are presented in Table~\ref{t:MS-FS}. The table presents IS between reference isotope with atomic number $A$ given in second column of the table and an isotope with $A^\prime = A+2$. 
It is assumed that $M=A$~amu, nuclear radius $R_N =1.1A^{1/3}$ is the radius of the Fermi distribution function. Nuclear root-mean square radius (RMS) is calculated as $\langle r^2\rangle = \int \rho r^2 dV$, where $\rho$ is the nuclear density normalised to 1: $\int \rho dV=1$. SMS is neglected, the total mass shift is calculated as $(m_e/$amu$)[1/A -1/(A+2)]\times E_k$.
Field shift is calculated using the values of $F$ and $G$ from table~\ref{t:FGMS} and the values of $\delta \langle r^2 \rangle$ calculated via the Fermi charge distribution. Note that the terms with $G$ gives negligible contribution for light atoms. We keep it for consistency since for heavy atoms its contribution is not so small. The last column of Table~\ref{t:MS-FS} present the ratio of the total normal mass shift to the total field shift. This ratio $\sim 1$ in the vicinity of $Z=38$.
For lighter atoms mass shift dominates while for heavier atoms field shift dominates. 

\begin{table}
\caption{\label{t:MS-FS} Comparison of  the  normal mass shift with the field shift for isotopes with $\Delta A=2$.
It is assumed that $R_N = 1.1 A^{1/3}$. All numbers are in atomic units.}
\begin{ruledtabular}
\begin{tabular}   {rrl cccc}
\multicolumn{1}{c}{$Z$}&
\multicolumn{1}{c}{$A$}&
\multicolumn{1}{c}{Atom}&
\multicolumn{1}{c}{NMS}&
\multicolumn{1}{c}{FS}&
\multicolumn{1}{c}{Sum}&
\multicolumn{1}{c}{NMS/FS}\\
\hline   % a.u.
  10 &  20 &  Ne   &  3.22[-4] &  1.70[-6] &  3.24[-4] &  1.89[+2] \\
  12 &  24 &  Mg   &  3.52[-4] &  3.45[-6] &  3.56[-4] &  1.02[+2] \\
  18 &  36 &  Ar   &  2.08[-4] &  1.69[-5] &  2.25[-4] &  1.23[+1] \\
  20 &  40 &  Ca   &  4.44[-4] &  2.52[-5] &  4.69[-4] &  1.76[+1] \\
  30 &  64 &  Zn   &  4.66[-4] &  1.32[-4] &  5.98[-4] &  3.53[+0] \\
  36 &  78 &  Kr   &  4.90[-4] &  2.93[-4] &  7.84[-4] &  1.67[+0] \\
  38 &  84 &  Sr   &  4.83[-4] &  3.73[-4] &  8.55[-4] &  1.29[+0] \\
  48 & 106 &  Cd   &  5.36[-4] &  1.13[-3] &  1.67[-3] &  4.74[-1] \\
  50 & 112 &  Sn   &  5.30[-4] &  1.34[-3] &  1.87[-3] &  3.95[-1] \\
  54 & 128 &  Xe   &  4.91[-4] &  2.01[-3] &  2.51[-3] &  2.44[-1] \\
  56 & 130 &  Ba   &  5.20[-4] &  2.40[-3] &  2.92[-3] &  2.16[-1] \\
  70 & 170 &  Yb   &  5.27[-4] &  8.59[-3] &  9.12[-3] &  6.13[-2] \\
  78 & 194 &  Pt   &  5.31[-4] &  1.78[-2] &  1.84[-2] &  2.98[-2] \\
  80 & 200 &  Hg   &  5.33[-4] &  2.06[-2] &  2.11[-2] &  2.59[-2] \\
  86 & 220 &  Rn   &  5.29[-4] &  3.45[-2] &  3.50[-2] &  1.53[-2] \\
 102 & 259 &  No   &  5.95[-4] &  1.47[-1] &  1.47[-1] &  4.06[-3] \\
 118 & 294 &  Og   &  6.89[-4] &  6.72[-1] &  6.73[-1] &  1.02[-3] \\
 120 & 300 &  E120 &  6.95[-4] &  8.26[-1] &  8.26[-1] &  8.41[-4] \\
 
%  10 &  20 &  Ne   &    0.0088 &  4.64[-5] &  8.80[-3] &  1.89[+2] \\  eV
%  12 &  24 &  Mg   &    0.0096 &  9.38[-5] &  9.68[-3] &  1.02[+2] \\
%  18 &  36 &  Ar   &    0.0057 &  4.59[-4] &  6.12[-3] &  1.23[+1] \\
%  20 &  40 &  Ca   &    0.0121 &  6.85[-4] &  1.28[-2] &  1.76[+1] \\
%  30 &  64 &  Zn   &    0.0127 &  3.59[-3] &  1.63[-2] &  3.53[+0] \\
%  36 &  78 &  Kr   &    0.0133 &  7.99[-3] &  2.13[-2] &  1.67[+0] \\
%  38 &  84 &  Sr   &    0.0131 &  1.01[-2] &  2.33[-2] &  1.29[+0] \\ % MS=FS
%  48 & 106 &  Cd   &    0.0146 &  3.07[-2] &  4.53[-2] &  4.74[-1] \\
%  50 & 112 &  Sn   &    0.0144 &  3.65[-2] &  5.09[-2] &  3.95[-1] \\
%  54 & 128 &  Xe   &    0.0134 &  5.48[-2] &  6.82[-2] &  2.44[-1] \\
%  56 & 130 &  Ba   &    0.0141 &  6.53[-2] &  7.95[-2] &  2.16[-1] \\
%  70 & 170 &  Yb   &    0.0143 &  2.34[-1] &  2.48[-1] &  6.13[-2] \\
%  78 & 194 &  Pt   &    0.0145 &  4.85[-1] &  4.99[-1] &  2.98[-2] \\
%  80 & 200 &  Hg   &    0.0145 &  5.60[-1] &  5.74[-1] &  2.59[-2] \\
%  86 & 220 &  Rn   &    0.0144 &  9.39[-1] &  9.54[-1] &  1.53[-2] \\
% 102 & 259 &  No   &    0.0162 &  3.99[+0] &  4.01[+0] &  4.06[-3] \\
% 118 & 300 &  Og   &    0.0180 &  1.82[+1] &  1.82[+1] &  9.91[-4] \\
% 120 & 305 &  E120 &    0.0183 &  2.23[+1] &  2.24[+1] &  8.19[-4] \\
\end{tabular}
%\footnotetext[1]{Ref.~\cite{dwn}.} 
\end{ruledtabular}
\end{table}

Therefore, for heavy atoms we calculate only FS constants $F$ and $G$. The results are presented in Table~\ref{t:FGA}. 
We include some open-shell atoms in the interval $112 < Z < 118$. This is useful to test the interpolation formula (see below).
The calculations for the open-shell atoms have been performed as for closed-shell atoms but with fractional occupation numbers.
Apart from IS given by (\ref{e:FG}), we also considered field  IS  fitted by the formula suggested in Ref.~\cite{IS13} to describe IS in superheavy elements (compare with Eq. (\ref{e:2gamma})), 
\begin{equation}\label{e:da}
\Delta E_{\rm IS} = a\left( A_1^{1/3} -  A_2^{1/3} \right).
\end{equation}
This formula may  be used when the change of the nuclear radius between the isotopes  is not known.
We calculate its value by changing nuclear radius and fitting the change of the total electron binding energy to (\ref{e:da}). 
The results are presented in last column of Table~\ref{t:FGA}.

\begin{table}
  \caption{\label{t:FGA}Field shift constants for heavy atoms. For $F$ and $G$ see formula
  (\ref{e:FG}); for $a$ see formula (\ref{e:da}).}
\begin{ruledtabular}
\begin{tabular}   {rl lccc}
\multicolumn{1}{c}{$Z$}&
\multicolumn{1}{c}{Atom}&
\multicolumn{1}{c}{GS}&
\multicolumn{1}{c}{$F$}&
\multicolumn{1}{c}{$G$}&
\multicolumn{1}{c}{$a$}\\
&&\multicolumn{1}{c}{conf.}&
\multicolumn{1}{c}{a.u./fm$^2$}&
\multicolumn{1}{c}{a.u./fm$^4$}&
\multicolumn{1}{c}{a.u.}\\
\hline  % a.u.
  48 & Cd & [Kr]$4d^{10}5s^2$  &  0.0055 &  -0.0000 &   0.0408 \\
%  50 & Sn   &  0.0067 &  -0.0000 &   0.0501 \\
  54 & Xe & [Cd]$5p^{6}$  &  0.0105 &  -0.0000 &   0.0815 \\
%  54 & Xe   &  0.1052E-01 &  -0.1875E-04 &   0.8155E-01 \\
  56 & Ba & [Xe]$6s^{2}$  &  0.0126 &  -0.0000 &   0.0982 \\ % 48-56 k=5.38

  70 & Yb  & [Ba]$4f^{14}$ &  0.0493 &  -0.0001 &   0.4178 \\  % 56-70 k=6.11
%  73 & Ta & [Yb]$5d^{3}$  &  0.0672 &  -0.0002 &   0.5778 \\
%  74 & W   & [Yb]$5d^{4}$ &  0.0735 &  -0.0002 &   0.6361 \\
%  75 & Re  & [Yb]$5d^{5}$ &  0.0809 &  -0.0002 &   0.6987 \\
%  76 & Os  & [Yb]$5d^{6}$ &  0.0887 &  -0.0003 &   0.7688 \\
%  77 & Ir  & [Yb]$5d^{7}$ &  0.0973 &  -0.0003 &   0.8438 \\
  78 & Pt  & [Xe]$4f^{14}5d^{10}$ &  0.1068 &  -0.0003 &   0.9293 \\ % - 5d9 6s
%  79 & Au  & [Pt]$6s$ &  0.1171 &  -0.0004 &   1.0227 \\  
  80 & Hg  & [Pt]$6s^{2}$ &  0.1246 &  -0.0004 &   1.0975 \\  % 80-70 k=6.94

  86 & Rn  & [Hg]$6p^{6}$ &  0.2157 &  -0.0008 &   1.9593 \\  %80-86 k=7.59
 102 & No & [Rn]$5f^{14}7s^2$  &  0.9679 &  -0.0050 &   9.0064 \\
% 110 & Ds  & [No]$6d^{8}$ &  2.1395 &  -0.0132 &  20.1935 \\
% 111 & Rg  & [No]$6d^{9}$ &  2.3635 &  -0.0153 &  22.2694 \\ % 0.22 2.08
 112 & Cn & [No]$6d^{10}$  &  2.5520 &  -0.0163 &  24.2184 \\ % 0.19 1.95 - 86-112 k=9.35
 113 & Nh  & [Cn]$7p$ &  2.8624 &  -0.0186 &  27.2553 \\ % 0.31 3.04
 114 & Fl  & [Cn]$7p^{2}$ &  3.1364 &  -0.0203 &  30.1942 \\
 115 & Mc & [Cn]$7p^{3}$  &  3.4693 &  -0.0230 &  33.3857 \\
 116 & Lv & [Cn]$7p^{4}$ &  3.8395 &  -0.0260 &  36.9329 \\
 117 & Ts  & [Cn]$7p^{5}$ &  4.2468 &  -0.0295 &  40.8796 \\
 118 & Og  & [Cn]$7p^{6}$ &  4.6288 &  -0.0330 &  44.6770 \\ % 112-118 k=11.41
% 119 & E119 & [Og]$8s$&  5.1816 &  -0.0378 &  50.1591 \\
 120 & E120 & [Og]$8s^{2}$&  5.7224 &  -0.0424 &  55.6019 \\ %  112-120 k=11.70
\end{tabular}
\end{ruledtabular}
\end{table}
%\begin{equation}\label{e:ak}
%F(Z) = aZ^k
%\end{equation}

\begin{table}
  \caption{\label{t:FGA1}Field shift constants for heavy atomic ions. See Table~\ref{t:FGA} for notations.}
\begin{ruledtabular}
\begin{tabular}   {rl lccc}
\multicolumn{1}{c}{$Z$}&
\multicolumn{1}{c}{Ion}&
\multicolumn{1}{c}{GS}&
\multicolumn{1}{c}{$F$}&
\multicolumn{1}{c}{$G$}&
\multicolumn{1}{c}{$a$}\\
&&\multicolumn{1}{c}{conf.}&
\multicolumn{1}{c}{a.u./fm$^2$}&
\multicolumn{1}{c}{a.u./fm$^4$}&
\multicolumn{1}{c}{a.u.}\\
\hline  % a.u.  49 & In   & [Cd] &  0.0062 &  -0.0000 &   0.0449 \\
  49 & In$^+$   & [Cd]   &  0.0062 &  -0.0000 &   0.0449 \\
  55 & Cs$^+$   & [Xe] &  0.0118 &  -0.0001 &   0.0891 \\
  57 & La$^+$   & [Ba] &  0.0145 &   -0.0001 &   0.1120 \\
  71 & Lu$^+$   & [Yb] &  0.0557 &  -0.0002 &   0.4677 \\
  79 & Au$^+$   & [Pt] &  0.1171 &  -0.0004 &   1.023 \\
  81 & Tl$^+$   & [Hg] &  0.1407 &  -0.0005 &   1.240 \\
  87 & Fr$^+$   & [Rn] &  0.2433 &  -0.0009 &   2.210 \\
 103 & Lr$^+$   & [No] &  1.088  &  -0.0058 &   10.16 \\
 113 & Nh$^+$   & [Cn] &  2.862  &  -0.0186 &   27.26 \\
 119 & E119$^+$ & [Og] &  5.181  &  -0.0377 &   50.16 \\
\end{tabular}
\end{ruledtabular}
\end{table}

To choose the formula for the interpolation of the FS constant $F$ for intermediate values of $Z$ we note that the dominating contribution comes from  the deep $s$ and $p_{1/2}$ shells where dependence on $Z$ in heavy atoms is very strong  - see e.g. Ref. \cite{gamma}. 
%Here FS shoes strong dependence on $Z$:
%\begin{equation}\label{e:1s}
%F \sim (A^{1/3}Z)^{2\gamma},
%\end{equation}
%where $\gamma=\sqrt{1-Z^2\alpha^2}$.
 We have found that an  interpolation between neighbours in Tables  ~\ref{t:FGA} and  ~\ref{t:FGA1} with the accuracy better than 1\%  may be achieved  using this approximation:
\begin{equation}\label{e:Fk}
F(Z) = bZ^k.
\end{equation}
Taking $F(Z_1) = F_1$ and $F(Z_2) = F_2$ we get 
\begin{equation}\label{e:Fk}
F(Z) = F_1\left(\frac{Z}{Z_1}\right)^k, \ \ k=\frac{\ln(F_1/F_2)}{\ln(Z_1/Z_2)}.
\end{equation}
The value of $k$ rapidly increases with $Z$,   from about 5 near $Z\sim 50$ to $12.6$ at $Z=118$ - 120, reflecting the rapidly increasing contribution of deep $s$ and $p_{1/2}$ shells.
 %and  for the superheavy elements  Z=118 and Z=120  reaches $k=12.6$. 
%for given $Z$-intervals between closed-shell atoms are presented in Table~\ref{t:ak}.
%Comparing $F(Z)$ given by (\ref{e:Fk}) with calculated values of $F$ for $112 < Z< 118$ (Table~\ref{t:FGA}) shows that the accuracy of the interpolation $\sim$~1\%. 
A similar interpolation may be done for the constants $G$.

\section{Conclusion}\label{sec:conclusion}
We have carried out a systematic  study of isotope shifts of the \emph{total} electron binding energy for  neutral atoms and singly charged ions up to $Z=120$. Using relativistic Hartree-Fock wave functions with the Breit interaction included, we extracted field shift coefficients by varying the nuclear radius and fitting the resulting energy changes with a linear-quadratic form,  $F\delta\langle r^2\rangle + G\delta\langle r^2\rangle^2$. Tables of $F$ and $G$ are provided for closed shell systems from Ne to Og, with benchmark calculations for several open-shell cases, used to test the interpolation formula. For heavy elements we also supply the alternative coefficient $a$ for a compact $A^{1/3}$ parametrization of the field  shift, useful when radius changes are not independently known.

A simple power law interpolation  $bZ^k$ reproduces calculated field shifts between neighboring closed shells to within $\sim 1\%$ across the periodic table; the effective exponent $k$ rises from about 5 near $Z\sim 50$ to $12.6$ at $Z=118$ - 120, reflecting the rapidly increasing contribution of deep $s$ and $p_{1/2}$ shells.

 Differences between neutrals and singly charged ions are at the few percent level and become noticeable mainly when an outer $s$ electron is removed.
Therefore, our results may also be used for ions with a higher charge.

Comparison of mass and field contributions shows a crossover near $Z \approx 38$: the mass shift dominates below and the field shift above this point, with an overwhelming field shift dominance in superheavy elements. 
%The specific mass shift contributes only a modest fraction of the normal mass shift (about 10% in light atoms, a little over 20% in Xe), and is negligible for practical applications in heavy systems.

Isotope shift in the total  electron energy  is dominated by the contribution of deep shells, therefore  correlation corrections are expected to be small. The uncertainties of the tabulated field shift factors are expected to be at the percent level for heavy atoms.
% limited chiefly by the nuclear?charge distribution model rather than electronic correlations, which are small for the deep shells that determine the total?energy shift. 

\end{document}